\documentclass[a4paper,12pt]{article}

\usepackage{lineno,hyperref}
\modulolinenumbers[5]%
\usepackage{hyperref}
\usepackage[authoryear]{natbib}

\usepackage{amsmath}

\begin{document}


\title{Relation between firing statistics of spiking neuron with instantaneous feedback and without feedback}

\author{Alexander K.Vidybida %
\thanks{Bogolyubov Institute for Theoretical Physics, Metrologichna str., 14-B, Kyiv 03680, Ukraine, vidybida@bitp.kiev.ua, http://www.bitp.kiev.ua/pers/vidybida}
}
\date{}


\maketitle
\thispagestyle{empty} 
\bibliographystyle{model2-names.bst}

\begin{abstract}
We consider a wide class of spiking neuron models, defined by rather general set of conditions
typical for basic models like leaky integrate and fire, or binding neuron model.
A neuron is fed with a point renewal process. 
A relation between the three probability density functions (pdf): (i) pdf of input interspike intervals,
(ii) pdf of output interspike intervals of a neuron with instantaneous feedback and 
(iii) pdf for that same neuron without feedback is derived. This allows to calculate 
any of the three pdfs provided the another two are given. Similar relation between corresponding
means and variances is derived. The relations are checked exactly for the binding neuron model.


MSC ~~ 60K40; 60K20; 92B20
\end{abstract}





\section{Introduction}

Spiking statistics of various neuronal models under a random stimulation is for a long
time considered in the framework of diffusion approximation, see \cite{Sacerdote2013}
and references therein. In the diffusion approximation, it is difficult to follow the 
fate of individual impulses, which is required for studying a neural network, 
or the simplest case of it --- a neuron with a feedback. Recently, a progress has been made
in description of neuronal firing statistics without using diffusion approximation,
see e.g. \citep{Arunachalam2013,Vidybida2007,Vidybida2008}. In the cited papers, output statistics
is calculated for binding neuron \citep{Vidybida2007} and binding neuron with instantaneous 
feedback \citep{Arunachalam2013,Vidybida2008} stimulated with Poisson point process and
some other processes \citep{Arunachalam2013}. In this letter we offer a simple relation
between output statistics of a neuron with instantaneous feedback and 
that same neuron without feedback,
which is valid for any renewal input point process. Also, we do not specify a concrete neuronal model,
only formulate a set of conditions the model must satisfy. Most basic neuronal models
do satisfy the formulated conditions.

\section{Assumptions and definitions}

The main function of a neuron is to transform its stream of input impulses (the stimulus)
into its stream of output impulses. An output impulse is usually called ``spike''.
When a neuron emits an output impulse, it is usually said that neuron is triggered and fires a spike. 
As regards neuronal functioning we assume that the following conditions are satisfied:

\begin{itemize}
\item[COND1] 
Neuron is stimulated with excitatory input impulses which form a renewal point process.
The process is described by means of a probability density function (pdf) of input interspike intervals (ISI), $p^{in}(t)$, where $t$ denotes an ISI duration.
\item[COND2] 
Neuron has a deterministic behavior: the same stimulus, which is a sequence of input impulses,
 gives the same result (the neuron either fires,  or does not fire).
\item[COND3] 
After firing, neuron appears in its resting state, which does not evolve in time
until a fresh input impulse comes.
\item[COND4]
Neuron may fire only at a moment when an input impulse comes.
\item[COND5]
If neuron starts from its resting state, then more than one input impulse is required
in order to trigger it.
\item[COND6]
Output stream of impulses can be characterized with pdf of the output ISIs, $p^{o}(t)$.
\end{itemize}

The conditions COND1-COND6, above, are satisfied for basic neuronal models, such as perfect
 integrate-and-fire
model, see \citep{Abbott1999}, leaky integrate-and-fire model \citep{Stein1967}, or binding neuron model \citep{Vidybida2007}.
Condition COND5 means that the neuron makes some processing of sets of input impulses, instead of
channelizing every input impulse into its output stream.

From COND1-COND4 it follows that the output stream of ISIs will be as well a renewal stochastic process.

The above construction can be extended by adding an instantaneous feedback line, see \citep{Vidybida2008}. The line sends any
output impulse to the neuronal input without delay. This impulse is identical to any other input
impulse. In this case we have a neuron with instantaneous feedback (IF).

It is worth noticing that immediately after firing the neuron with IF as well appears in a
standard initial state. This standard initial state is realized if a neuron being in the resting state
gets one input impulse. This state can evolve in time\footnote{This state evolves with time for any
 neuronal model (received with impulse excitation decays in time), except of the perfect integrator.}. 
Nevertheless, it is clear that the output stream of ISIs
of a neuron with IF as well will be a renewal process. As regards this process, we expect the 
following:
\begin{itemize}
\item[COND7]
The output stream of ISIs of neuron with IF  can be described by means of a pdf $p^{o\_if}(t)$,
where "if" stands for the "instantaneous feedback".
\end{itemize}

\section{Relation between pdfs}

Our purpose is to establish a relation between $p^{in}(t)$, $p^{o}(t)$ and $p^{o\_if}(t)$.
Any of the three processes is a renewal one. If so, it is enough to analyze what happens between two
consecutive firings of neuron without feedback. 

Expect that neuron without feedback fires at moment 0. In order to fire next time at moment from
$[t;t+dt[$, the neuron must obtain an input impulse at this same moment (COND4), and this impulse
is not the first one received after the moment 0 (COND5). The first one must be obtained earlier, at some 
moment $t'\in ]0;t[$. The probability to receive this impulse in the interval $[t';t'+dt'[$
is $p^{in}(t')dt'$. After receiving this impulse, the neuron appears in the standard initial state
of neuron with instantaneous feedback. Now, firing next time at $t$ means that a neuron with IF
fires first time at $t$ if starts at $t'$. This event does not depend on the event of 
receiving first input impulse and has probability $p^{o\_if}(t-t')dt$. Thus, the compound event
of receiving the first impulse at time $t'<t$ and firing firstly at time $t$
has the following probability
$
p^{in}(t')dt'p^{o\_if}(t-t')dt.
$
This expression gives probability of a single alternative, where the whole
set of alternatives is parameterized with $t'$. To get the $p^{o}(t)dt$ one has to add the probabilities of all alternatives:
\begin{equation}\label{addalter}
p^{o}(t)dt=\int\limits_0^t p^{in}(t') p^{o\_if}(t-t')dt'dt.
\end{equation}
The required relation follows from (\ref{addalter}):
\begin{equation}\label{rel}
p^{o}(t) = \int\limits_0^t p^{in}(t')p^{o\_if}(t-t')dt'.
\end{equation}

\subsection{Inverting Eq. (\ref{rel})}
In general case, the Eq. (\ref{rel}) can be inverted by means of Laplace transform:
$$
\mathcal{L}(p^{o})(s)=\mathcal{L}(p^{in})(s)\mathcal{L}(p^{o\_if})(s),
$$
\begin{equation}\label{Laplace}
\mathcal{L}(p^{o\_if})(s)={\mathcal{L}(p^{o})(s)}/{\mathcal{L}(p^{in})(s)}.
\end{equation}
In order to find exact expression for $p^{o\_if}(t)$ it is necessary to apply the inverse Laplace
transform to the right hand side of (\ref{Laplace}). This operation can be accomplished depending on the
explicit expressions for the $p^{in}(t)$, $p^{o}(t)$.

\subsection{Poissonian input case}
In this case $p^{in}(t) = \lambda e^{-\lambda t}$ and its Laplace transform is
$
\mathcal{L}(p^{in})(s) = \frac{\lambda}{s+\lambda},
$
which gives after using in (\ref{Laplace})
\begin{equation}\label{relPois}
\mathcal{L}(p^{o\_if})(s)=\mathcal{L}(p^{o})(s)+s\mathcal{L}(p^{o})(s)/\lambda.
\end{equation}
Notice that from COND1, COND5 it follows that $p^{o}(0)=0$. If so, then (\ref{relPois}) gives
\begin{equation}\label{relPoisdiff}
p^{o\_if}(t) = p^{o}(t) + \frac{1}{\lambda}\frac{d}{dt}p^{o}(t).
\end{equation}

\section{Example - binding neuron with threshold two}

The binding neuron (BN) model is characterized with a time interval $\tau>0$ during 
which an input impulse is stored in the neuron. The BN with threshold 2
fires a spike at the moment of receiving an input impulse, provided at that moment
the previous input impulse is still stored in the neuron. Just after firing, BN is free
of stored impulses.
If BN is stimulated with the Poisson stream, then COND1-COND7 are satisfied.

Exact expression for the $p^{o}(t)$ can be found in \citep[Eq. (3)]{Vidybida2007}:
for $m=0,1,2,\ldots$, if $m\tau<t\le (m+1)\tau$ then
\begin{multline}\label{exact}
p^{o}(t)=
e^{-\lambda t}\frac{\lambda^{m+2}}{(m+1)!}(t-m\tau)^{m+1}
\\
+
e^{-\lambda t}
\sum\limits_{2\le k\le m+1}
\frac{\lambda^k}{(k-1)!}
\left((t-(k-2)\tau)^{k-1} - (t-(k-1)\tau)^{k-1}\right).
\end{multline}

Exact expression for the $p^{o\_if}(t)$ can be found in \citep[Eqs. (4),(7)]{Vidybida2008},
or in \citep[Eq. (10)]{Arunachalam2013}:
for $m=0,1,2,\ldots$, if $m\tau<t\le (m+1)\tau$ then
\begin{multline}\label{ifexact2}
p^{o\_if}(t)=
e^{-\lambda t}{\lambda^{m+1}\over m!}(t-m\tau)^m
\\+
e^{-\lambda t}
\sum\limits_{2\le k\le m}
{\lambda^k\over(k-1)!}
\left((t-(k-1)\tau)^{k-1} - (t-k\,\tau)^{k-1}\right)\,.
\end{multline}
Now, the validity of (\ref{relPoisdiff}) for BN can be directly checked by substituting
(\ref{exact}), (\ref{ifexact2}) into it.

\section{Moments of distribution}

Denote $W^{\{in,o,o\_if\}}_n$ the $n$-th moment of the corresponding distribution.
With using (\ref{rel}) one has
\begin{multline}\nonumber
W^o_n=\int\limits_0^\infty dt\,t^n\, p^{o}(t)
=\int\limits_0^\infty dt\,t^n\int\limits_0^\infty dt'p^{in}(t')p^{o\_if}(t-t')
\\
=\int\limits_0^\infty dt'p^{in}(t')\int\limits_{t'}^\infty dt\,t^n\, p^{o\_if}(t-t')
\\
=\int\limits_0^\infty dt'p^{in}(t')\int\limits_{0}^\infty dt\,(t+t')^n p^{o\_if}(t)
=\sum\limits_{k=0}^n \dbinom{n}{k}W_k^{in}W_{n-k}^{o\_if}\,.
\end{multline}
In particular, for $n=1,2$ one has
\begin{equation}\label{W1}
W_1^o=W_1^{o\_if}+W_1^{in},
\end{equation}
\begin{equation}\label{W2}
W_2^o=W_2^{o\_if}+W_2^{in}+2W_1^{in}W_1^{o\_if}.
\end{equation}

For binding neuron with threshold 2 fed with Poisson stream, (\ref{W1}) can be checked explicitly.
It is clear that here $W_1^{in}=\frac{1}{\lambda}$. The $W_1^o$ is found in \citep[Sec. 2]{Vidybida2007}:
$$
W_1^o=\frac{1}{\lambda}\left(2+\frac{1}{e^{\lambda\tau} -1}\right).
$$
The $W_1^{o\_if}$ is found in \citep[Eq. (9)]{Vidybida2008} or \citep[Eq. (11)]{Arunachalam2013}:
$$
W_1^{o\_if}=\frac{1}{\lambda\left(1-e^{-\lambda\tau}\right)}.
$$
The validity of (\ref{W1}) can now be checked by substituting these expressions into it.
Similarly, \citep[Eqs. (12), (13)]{Vidybida2008} give:
$$
W_2^o=\frac{2}{\lambda^2}\,\frac{3\,e^{2\,\lambda\,\tau}+(\lambda\,\tau-3)\,e^{\lambda\,\tau}+1}{(e^{\lambda\,\tau}-1)^2},\quad
W_2^{o\_if}=\frac{2\,e^{\lambda\,\tau}}{\lambda^2}\,\frac{e^{\lambda\,\tau}+\lambda\,\tau}{\left(e^{\lambda\,\tau}-1\right)^2}\,.
$$
The validity of (\ref{W2}) for binding neuron with threshold 2 fed with Poisson stream
can be checked by substituting these expressions into it.

Finally, denote $\sigma^2_{\{in,o,o\_if\}}$ the variance of corresponding distribution.
Then from (\ref{W1}), (\ref{W2}) the following relation can be derived:
\begin{equation}
\sigma^2_{o\_if} = \sigma^2_{o} - \sigma^2_{in}.
\end{equation}

\bigskip

{\small\bf Acknowledgements.}
{\small This paper was supported by the Program
"Microscopic and phenomenological models of fundamental physical
processes in a micro and macroworld" of the National Academy of Science of Ukraine, Project PK No 0112U000056.}



\bibliography{../References}{}

\end{document}